\documentclass[prd,aps,twocolumn,showpacs,preprintnumbers,amsmath,amssymb,nofootinbib]{revtex4}
\usepackage[dvips]{graphicx}
\usepackage{epsf}
\usepackage{amsmath}
\usepackage{amssymb}

\usepackage{graphicx}
\usepackage{dcolumn}
\usepackage{bm}
\def\be{\begin{equation}}
\def\ee{\end{equation}}
\def\bea{\begin{eqnarray}}
\def\eea{\end{eqnarray}}

\usepackage{color}

\newcommand{\comment}[1]{}

\begin{document}


\title{The Matter Bounce Curvaton Scenario}

\author{Yi-Fu Cai$^{a,b}$, Robert Brandenberger$^{c,b}$ and Xinmin Zhang$^{b,d}$}

\affiliation{a Department of Physics, Arizona State University,
Tempe, AZ 85287, USA}

\affiliation{b Institute of High Energy Physics, Chinese Academy
of Sciences, P.O. Box 918-4, Beijing 100049, P.R. China}

\affiliation{c Department of Physics, McGill University,
Montr\'eal, QC, H3A 2T8, Canada}

\affiliation{d Theoretical Physics Center for Science Facilities
(TPCSF), Chinese Academy of Sciences, P.R. China}


\begin{abstract}

Massless scalar fields originating in a quantum vacuum state acquire a scale-invariant spectrum of fluctuations in a matter-dominated contracting universe. We show that these isocurvature fluctuations transfer to a scale-invariant spectrum of curvature fluctuations during a non-singular bounce phase. This provides a mechanism for enhancing the primordial adiabatic fluctuations and suppressing the ratio of tensor to scalar perturbations. Moreover, this mechanism leads to new sources of non-Gaussianity of curvature perturbations.

\end{abstract}

\maketitle

\newcommand{\eq}[2]{\begin{equation}\label{#1}{#2}\end{equation}}

\section{Introduction}

A successful resolution of the hot Big Bang singularity may lead to a non-singular bounce cosmology. Non-singular bounces were proposed a long time ago \cite{Tolman:1931zz}. They were studied in models motivated by approaches to quantum gravity such as modified gravity models \cite{Brustein:1997cv, Cartier:1999vk}, higher derivative gravity actions (see e.g. \cite{Mukhanov:1991zn, Tsujikawa:2002qc, Biswas:2005qr}), non-relativistic gravitational actions \cite{Brandenberger:2009yt, Cai:2009in, Suyama:2009vy, Gao:2009wn}, brane world scenarios \cite{Kehagias:1999vr, Shtanov:2002mb},
``Pre-Big-Bang" \cite{PBB} and Ekpyrotic \cite{Ekp} cosmology \footnote{Some versions are non-singular,
see e.g. \cite{PBBns} in the case of the Pre-Big-Bang scenario, or the ``New Ekpyrotic scenario"
\cite{Khoury,Creminelli}.}, or loop quantum cosmology \cite{Bojowald:2001xe}. Bouncing cosmologies can also be obtained using arguments from superstring theory. For example, the string gas cosmology scenario \cite{Brandenberger:1988aj, Brandenberger:2008nx} may be embedded in a bouncing universe as realized in \cite{Biswas:2006bs}. Non-singular bounces may also be studied using effective field theory techniques by introducing matter fields violating certain energy conditions, for example non-conventional fluids \cite{Bozza:2005wn, Peter:2002cn}, quintom matter \cite{Cai:2007qw, Cai:2007zv, Cai:2008qb, Cai:2008ed, Cai:2009zp}, or ghost condensates \cite{Khoury,Creminelli,Chunshan}. A specific realization of  a quintom bounce occurs in the Lee-Wick cosmology studied in \cite{Cai:2008qw}. Various original bounce models were reviewed in Ref. \cite{Novello:2008ra}.

In the context of studies of bouncing cosmologies it has been realized that fluctuations which are generated as quantum vacuum perturbations and exit the Hubble radius during a matter-dominated contracting phase lead to a scale-invariant spectrum of cosmological fluctuations today \cite{Wands:1998yp, Finelli:2001sr} (see also \cite{Starobinsky:1979ty} for an earlier discussion). This yields an alternative to inflation for explaining the current observational data, which is dubbed as the {\it matter bounce} (see e.g. \cite{RHBrev} for a recent review). However, in this scenario there are still unclear issues. For example, since scalar and tensor modes grow at the same rate in a matter-dominated phase of contraction, the tensor-to-scalar ratio is typically too large to be consistent with current observational data. Also, the possible generation of entropy during the non-singular bounce phase needs to be studied in detail. In this paper, we will focus on the first question, leaving the second for a followup paper \cite{inprep}.

To date, entropy fluctuations have not been considered in matter bounce models. However, realistic matter theories will contain many entropy modes. In the context of inflationary cosmology, it has been realized that entropy fluctuations can lead to an additional source of curvature fluctuations, a source which may in fact dominate \cite{Mollerach, Linde, Lyth, Moroi, Sloth}. This is now know as the ``curvaton mechanism" for generating fluctuations. Entropy fluctuations also play an important role in the Ekpyrotic scenario \cite{Finelli, Notari, Turok, Creminelli, Khoury}. In that scenario, the primordial adiabatic fluctuations in the contracting phase acquire a steep blue spectrum, whereas the entropy modes induced by light scalar fields will be scale-invariant.

In this paper, we will consider the fluctuations induced by a light spectator field $\chi$ and see that they acquire a scale-invariant spectrum of fluctuations in a matter-dominated phase of contraction if they originate in their vacuum state on sub-Hubble scales. We will show that these entropy fluctuations induce a scale-invariant spectrum of curvature perturbations after the bounce. The transfer of entropy to adiabatic perturbations occurs during the bounce phase. We can easily find parameter values for which these secondary curvature perturbations are larger in magnitude than the primordial ones. This will lead to a suppression of the tensor to scalar ratio.

As a simple example, we consider a two-field model with potential
\begin{eqnarray}\label{potential}
 V(\phi, \chi) \, = \, \frac{1}{2}m^2\phi^2 +\frac{1}{2}g^2\phi^2\chi^2~,
\end{eqnarray}
where in our model $\phi$ is the background scalar field in the nonsingular bounce cosmology, and $\chi$ is a second scalar matter field, the so-called entropy field \footnote{The non-singular bounce is induced by adding another matter field violating the null energy condition to the system, or by considering the matter Lagrangian containing the regular fields $\phi$ and $\chi$ in the presence of a gravitational action leading to a bounce. Note that in a successful bouncing cosmology the usual energy conditions need to be recovered after the bounce so that the universe is able to evolve smoothly into the radiation and matter dominated periods. This feature exactly coincides with what occurs in the quintom dark energy model\cite{Feng:2004ad}, in which the dark energy equation of state crosses the cosmological constant boundary.}.

In the context of a matter bounce background, we study the cosmological evolution for both background and perturbations. We study the specific model given by the potential (\ref{potential}). We find that the entropy field $\chi$ evolves as a tracking solution in the contracting phase due to the coupling to the background scalar $\phi$, and may become dominant during a short deflationary phase before the bounce. The fluctuations of the entropy field are nearly scale-invariant provided the coupling parameter $g$ is small enough. A controllable amplification on the entropy modes due to the kinetic term of the entropy field can be obtained in the bouncing phase, and these entropy modes will be converted to curvature perturbations during the bounce. Therefore, we can obtain an enhancement of the primordial curvature spectrum and hence a suppression of the tensor-to-scalar ratio. This is the main message of this work. Another interesting fact is that the conversion of entropy fluctuations to curvature perturbation provides a sizable non-Gaussianity for curvature perturbations for which the sign tends to be negative.

The paper is organized as follows. In section II we provide the basic equations of the theory. In section III we study the generation of isocurvature fluctuations in the contracting phase. In section IV we analyze the conversion of the isocurvature perturbations into curvature fluctuations during the bounce and show that a reasonable amplitude of the primordial power spectrum can be obtained. Section V presents a summary and discussions of related works.

We use the convention $m_{pl}=1/\sqrt{G}$ in this paper.

\section{Setup and Basic Equations}

\begin{figure}[htbp]
\includegraphics[scale=0.3]{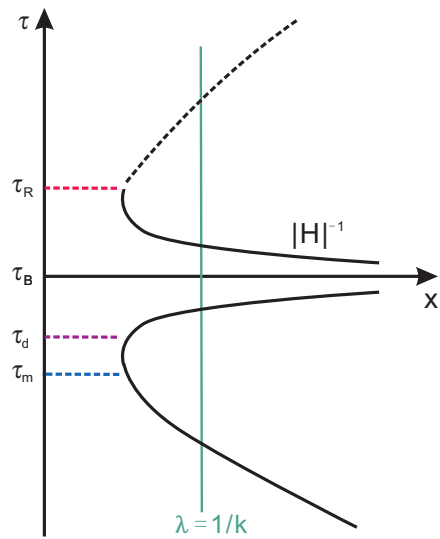}
\caption{A sketch of the evolution of scales in a nonsingular bouncing universe. The horizontal axis is comoving spatial coordinate, the vertical axis is conformal time. Plotted are the Hubble radius $|{ H}|^{-1}$ and the wavelength $\lambda$ of a fluctuation with comoving wavenumber $k$. } \label{Fig-st}
\end{figure}

To be specific, we will be considering the background space-time which emerges from a quintom bounce (see for example \cite{Cai:2008qw}) scenario in which regular matter is modelled by a scalar field $\phi$ with mass $m$ and the non-singular bounce is induced by adding a ghost scalar field $\psi_g$ with a much larger mass $M$ to the system. Fig. \ref{Fig-st} is a sketch of the evolution of scales in our matter bounce model. The vertical axis represents conformal time, the horizontal axis comoving distance. Initially, both the regular matter field and the ghost field are oscillating, with amplitudes which are increasing since the universe is contracting. During this period, the Hubble radius is decreasing linearly and $\dot{H} < 0$. We assume that the energy density is dominated by the regular field $\phi$. Once the amplitude of the regular field $\phi$ reaches the value $m_{pl}$, the field will cease to oscillate. This happens at conformal time $\tau_m$. A short deflationary ``slow climb'' phase will begin during which the Hubble radius $|H|^{-1}$ is nearly constant. In this period the
ghost field continues to oscillate. Hence, the (negative) energy density of the ghost field catches up to the (positive) energy density of the field $\phi$ and a short non-singular bounce period will result during which $\dot{H} > 0$. The conformal time $\tau_d$ is the transition between the deflationary phase and the bounce period.
After the bouncing phase there is a short period of inflation lasting until $\tau_R$, after which the universe enters the expanding phase with a normal thermal history.
We denote the bounce time by $\tau_B$.

In Fig. \ref{Fig-st} we plot the evolution of the length corresponding to a fixed comoving scale. This scale is the wavelength of the fluctuation mode $k$ ($k$ standing for the comoving wavenumber) which we want to follow. The wavelength begins in the contracting phase inside the Hubble radius, exits the Hubble radius during this phase
and re-enters the Hubble radius during the normal expanding phase.

It is  important that scales of interest exit the Hubble radius in the matter-dominated phase of contraction. This is required to obtain a scale-invariant spectrum of cosmological perturbations. Important is also that there is a phase during which the energy density in the entropy field grows relative to the energy density in the field $\phi$. This condition is easily satisfied in the quintom bounce scenario. It is required in order that the conversion of the entropy fluctuations to curvature perturbations is efficient \footnote{This is similar to the requirement in inflationary cosmology for the curvaton mechanism to be efficient.}. The efficiency of our transfer mechanism will depend on whether the initial energy density of the entropy field $\chi$ is larger than the (absolute value of) the energy density in the ghost field $\psi_g$. We shall see that our mechanism is more efficient if $\chi$ dominates \footnote{It would be of great interest to study to what extent the conversion of entropy fluctuations to curvature perturbations is efficient in other bounce models such as the ghost condensate scenario of \cite{Chunshan} which is (unlike the quintom bounce \cite{Johanna}) free of ghosts at the perturbative level, and stable against the addition of radiation and anisotropic stress.}.

The key goal of our paper is to understand how entropy perturbations are generated in the contracting phase, then pass through the bounce, and finally convert into curvature perturbations after the bounce. In the following, we will study cosmological perturbation theory in the nonsingular bouncing cosmological background described above.

We shall work in longitudinal gauge in which the linearized scalar metric fluctuations appear in the metric in the following way:
\begin{eqnarray}
 ds^2 \, = \, (1+2\Phi)dt^2-a^2(t)(1-2\Psi)d\vec{x}^2~,
\end{eqnarray}
where $t$ is cosmic time and $x^i$ are the comoving spatial coordinates. The scalar metric fluctuations are characterized by two functions $\Phi$ and $\Psi$ which depend both on space and time. We take matter to consist of a set of scalar fields $\phi_i$. If the gravitational action is the usual one, then in the absence of linearized anisotropic stress \footnote{This will be the case if matter is described by a set of scalar fields.} the off-diagonal components of the perturbative Einstein equations imply $\Psi=\Phi$ (see \cite{Mukhanov:1990me} for a comprehensive discussion of the theory of cosmological perturbations). By expanding the Einstein and matter equations to first order, we obtain the following perturbation equations \cite{Sasaki:1995aw, Langlois:1999dw, Gordon:2000hv}:
\begin{eqnarray}
\delta\ddot\phi_i &+& 3H\delta\dot\phi_i + [-\frac{\nabla^2}{a^2}\delta\phi_i
+ \sum_jV_{,ij}\delta\phi_j] \nonumber\\
&=& 4\dot\phi_i\dot\Phi-2V_{,i}\Phi~, \\
-3H\dot\Phi &+& (\frac{\nabla^2}{a^2}-3H^2)\Phi \nonumber\\
&=& 4\pi{G}\sum_i[\dot\phi_i\delta\dot\phi_i-\dot\phi_i^2\Phi+V_{,i}\delta\phi_i]~,\\
\dot\Phi + H\Phi  &=& 4\pi{G}\sum_i\dot\phi_i\delta\phi_i~,
\end{eqnarray}
where $H=\dot a/a$ is the Hubble parameter, and $V_{,i}$ denotes the derivative of the scalar field potential with respect to $\phi_i$.

We can recast the above equations in terms of the Sasaki-Mukhanov variables \cite{Sasaki:1986hm, Mukhanov:1988jd} which are defined as
\begin{eqnarray}
Q_i\equiv\delta\phi_i+\frac{\dot\phi_i}{H}\Phi~,
\end{eqnarray}
and in terms of which the equations of motion are given by
\begin{equation}\label{pertQ}
\ddot{Q}_i +3H\dot{Q}_i -\frac{\nabla^2}{a^2}Q_i
+\sum_j[V_{,ij}-\frac{8\pi{G}}{a^3}(\frac{a^3}{H}\dot\phi_i\dot\phi_j)^{.}]Q_j
\, = \, 0~.
\end{equation}

To combine the above equations, we usually define the quantity $\zeta$ which is the curvature perturbation on the uniform density slice, and which is given by
\begin{eqnarray}\label{zeta}
\zeta \, = \, H\frac{\sum_i\dot\phi_iQ_i}{\sum_j\dot\phi_j^2}~ .
\end{eqnarray}
This quantity is conserved on super-Hubble scales in an expanding universe if there are only adiabatic fluctuations. However, the presence of entropy fluctuations on large scales will lead to a growth of $\zeta$ which corresponds to the seeding of an adiabatic fluctuation mode by the entropy mode.

At linear order, the general equation for the time derivative of $\zeta$ is (see Appendix)
\begin{eqnarray}\label{dotzeta}
 \dot\zeta \, = \, -\frac{H}{\dot{H}}\nabla^2\Phi  -
 4\pi GH\sum_i\frac{Q_i}{\dot\phi_i}(\frac{\dot\phi_i^2}{\dot{H}})^{.}~.
\end{eqnarray}
On large scales, the first term of the r.h.s of Eq. (\ref{dotzeta}) is negligible.
In a two field model, this equation reduces to the following expression for the time derivative of $\zeta$ on large scales:
\begin{eqnarray}\label{dotzeta1}
(1+w)\dot\zeta=4\pi{G}\frac{\dot\phi^2+\dot\chi^2}{3H}
[\frac{Q_{\phi}}{\dot\phi}-\frac{Q_{\chi}}{\dot\chi}]
\frac{d}{dt}(\frac{\dot\phi^2-\dot\chi^2}{\dot\phi^2+\dot\chi^2})~,
\end{eqnarray}
where $w=p/\rho$ is the equation of state, with $p$ and $\rho$ being pressure and energy density respectively. Note, however, that in our case we have three fields, the regular matter field $\phi$, the isocurvature field $\chi$ and the ghost matter field
$\psi_g$ which is responsible for yielding the non-singular bounce.

In the following we will study how in our non-singular bouncing cosmology the presence of primordial isocurvature fluctuations induces via (\ref{dotzeta}) a contribution to the curvature perturbations. We will consider two cases: the first where the isocurvature field $\chi$ initially contributes more to the density than the ghost field, the second case where $\chi$ is sub-dominant at the background level to the ghost field.

\section{Isocurvature Perturbations in the Contracting Phase}\label{sec:iso}

In this section we show that the isocurvature mode acquires a scale-invariant spectrum of fluctuations on scales which exit the Hubble radius in the matter-dominated phase of contraction, provided the initial conditions are taken to be vacuum. We will consider the model of (\ref{potential}), where the entropy mode corresponds to the scalar field $\chi$ which is taken to be massless. Note that since the ghost field $\psi_g$ has a large mass, its spectrum will not be scale-invariant. It will be highly suppressed in the infrared. Hence, $\psi_g$ cannot play the role of the entropy field for our purpose.

We consider a matter bounce background evolution of a FRW universe in which the matter-dominated phase of contraction is realized by the oscillations of a massive scalar field $\phi$ around it vacuum state (see Refs. \cite{Wands:1998yp, Finelli:2001sr, Allen:2004vz, Peter:2008qz} for other realizations). Hence, the Hubble parameter can be approximately expressed by
\begin{eqnarray}\label{hubble}
 \langle H(t) \rangle \, = \, \frac{2}{3t}~,
\end{eqnarray}
where the angular brackets stand for averaging over time. The solution for the scalar field $\phi$ can be asymptotically expressed as
\begin{eqnarray}\label{phi}
\phi(t) \, \simeq \, \tilde\phi(t)\sin mt~,
\end{eqnarray}
with a time dependent amplitude,
\begin{eqnarray}\label{tphi}
\tilde\phi(t) \, \equiv \, \frac{m_{pl}}{\sqrt{3\pi}m|t|}~,
\end{eqnarray}
which yields an equation of state which has vanishing pressure after averaging over an oscillation period of the field.

Since the entropy field $\chi$ is much lighter than the background scalar $\phi$ and we assume that it does not have a large initial homogeneous mode,  it has nothing to do with the background evolution before the bounce phase. Its evolution is determined by the equation of motion
\begin{eqnarray}\label{eomchi}
 \ddot\chi + 3H\dot\chi + \frac{1}{2}g^2\tilde\phi^2\chi \, \simeq \, 0~,
\end{eqnarray}
where we have averaged the coherent oscillations of the background scalar in the mass term and inserted a factor $1/2$ instead (parametric resonance instabilities resulting as a consequence of the oscillatory nature of $\phi$ \cite{Traschen:1990sw, Kofman:1994rk, Shtanov:1994ce, Kofman:1997yn} - see \cite{Allahverdi:2010xz} for a recent review - will be studied in a subsequent paper). This equation yields the following solution for $\chi$:
\begin{eqnarray}\label{chi}
 \chi(t) \, \propto \, t^{-1+\frac{g^2m_{pl}^2}{3\pi m^2}}~,
\end{eqnarray}
which evolves at the same rate as the amplitude of the background scalar if $gm_{pl}\ll m$. In this case it is convenient to define a parameter $d_1 \equiv \frac{\chi_i}{\phi_i}$ (where the subscripts ``i'' indicate the initial value of the corresponding fields), and then in the matter-dominated contracting phase we have
$\chi \simeq d_1\phi$.

Let us now turn to the fluctuations of $\chi$. Let us recall \cite{Wands:1998yp, Finelli:2001sr} that the finding that curvature fluctuations acquire a scale-invariant spectrum in the matter bounce scenario is based on assuming that the canonically normalized curvature fluctuation variable starts out in its vacuum state early in the contracting phase. It is hence reasonable to assume that also the fluctuations of $\chi$ start out in their vacuum state on scales smaller than the Hubble radius early in the contracting phase. We focus on scale which exit the Hubble radius in the matter-dominated phase of contraction and will show that due to the growth of fluctuations on super-Hubble scales a flat spectrum of isocurvature perturbations results \footnote{As discussed in \cite{Hongli}, the spectrum has a break at the scale which exits the Hubble radius at the time of the transition between the matter-dominated phase of contraction and the following phase (most likely a radiation-dominated phase). On smaller scales the spectrum will be blue.}.

We are interested in the scalar perturbation $\delta\chi$, where we write
\begin{eqnarray}
 \chi(t,\vec{x}) \, = \, \chi(t) + \delta\chi(t,\vec{x})~.
\end{eqnarray}
It is important to focus on the canonically normalized perturbation variable $v=a\delta\chi$ in terms of which quantum vacuum initial conditions can be imposed. We work on the spatially-flat slice where the metric fluctuation $\Phi$ vanishes. On this slice, the scalar field perturbations satisfy the Klein-Gordon equation\cite{Wands:2008tv}
\begin{eqnarray}\label{eqv}
 v_k'' + (k^2 + a^2m_{\chi}^2 - \frac{a''}{a})v_k \, = \, 0~,
\end{eqnarray}
where we have defined the effective mass term of the entropy field $m_{\chi}^2 \, = \, V_{,\chi\chi}$ and made use of the conformal time $d\eta = adt$. A prime indicates a derivative with respect to $\eta$. Comparing the above equation of motion with the basic perturbation equation (\ref{pertQ}), one may notice that the effective mass originating from the kinetic term of $\chi$ and the coupling between the isocurvaure and adiabatic perturbations has been neglected. This is justified since it is subdominant in the matter-dominated contracting phase.

Initially, the $k^2$ term dominates Eq. (\ref{eqv}), and so we can neglect the mass term and gravitational term. Thus, the dynamics corresponds to a free scalar propagating in a flat spacetime, and the initial condition takes the form of the Bunch-Davies vacuum:
\begin{eqnarray}
 v_k \, \simeq \, \frac{e^{-ik\tau}}{\sqrt{2k}}~.
\end{eqnarray}
During the phase of background contraction, the quantum fluctuations exit the Hubble radius and become classical perturbations (see \cite{Martineau,Polarski} for an analysis of the classicalization). For $a\propto\tau^2$ and $|m_\chi^2|\ll H^2$, Eq. (\ref{eqv}) becomes
\begin{eqnarray} \label{squeeze}
 v_k'' + (k^2-\frac{2}{\tau^2})v_k \, \simeq \, 0~.
\end{eqnarray}
The second term inside the parentheses on the left hand side of this equation is the term which leads to the squeezing of the fluctuations, and the coefficient and time dependence is exactly that required to transform a blue vacuum spectrum into a scale-invariant spectrum \cite{Wands:1998yp,Finelli:2001sr}.

We can see this explicitly as follows: Making use of the vacuum initial condition, we obtain an exact solution to (\ref{squeeze}):
\begin{eqnarray} \label{sol}
 v_k \, = \, \frac{e^{-ik\tau}}{\sqrt{2k}}(1-\frac{i}{k\tau})~.
\end{eqnarray}
Note that, in contracting phase these perturbations are able to grow outside the Hubble radius. Their amplitudes will keep increasing until the universe enters the contracting phase ends. Thus, from (\ref{sol}) it follows that on scales larger than the Hubble
radius, the amplitude of the spectrum of $\delta\chi$ in the contracting phase is given by
\begin{eqnarray}\label{dchic}
 \delta\chi(t) \, = \, P_{\chi}^{\frac{1}{2}}(t) \, = \, \frac{|H(t)|}{4\pi}~,
\end{eqnarray}
which is a function of cosmic time. From Eq. (\ref{eqv}), we are able to calculate the spectral tilt $n_{\chi}$ of the primordial perturbations
\begin{eqnarray}\label{n sigma}
n_{\chi} \, \equiv \, \frac{d\ln P_{\chi}}{d\ln k} \, = \, \frac{2m_{\chi}^2}{3H^2}~,
\end{eqnarray}
where $m_{\chi}$ is the effective mass of the $\chi$ field induced by the coupling to $\phi$. In the specific model given by (\ref{potential}), we apply the amplitude of the background scalar $\tilde\phi$ and the ratio $d$, and then obtain the spectral tilt
\begin{eqnarray}\label{nchi}
n_{\chi} \, \simeq \, \frac{g^2m_{pl}^2}{2\pi{m}^2}~.
\end{eqnarray}
From this result, we find that in order to ensure that the isocurvature perturbation are nearly scale-invariant, one has to require $|gm_{pl}| \ll \sqrt{2\pi}m$.

One can also estimate the amplitude of tensor perturbations $h$. It is well known that the tensor fluctuation amplitude satisfies the same equation of motion as a free scalar field in the given cosmological background. Hence, following the analysis done above for the fluctuations in the scalar field $\chi$, we obtain a scale-invariant spectrum of tensor modes (again assuming that they start out in their vacuum state), and an amplitude of the spectrum given by
\begin{eqnarray}
h \, \sim \, H/m_{pl} \, .
\end{eqnarray}
The tensor perturbation amplitude has the same order of magnitude as that of the adiabatic scalar metric perturbations. Thus (in terms of the usual ratio $r\equiv P_T/P_{\zeta}$ of the tensor to scalar metric power spectrum), the matter bounce
usually leads to a dangerously large value of $r$. From current observational CMB data \cite{Komatsu:2010fb}, this ratio is required to be less than $0.2$. Consequently, in order that a matter bounce scenario be consistent with current observations on the value of $r$, it is necessary that there be a mechanism to amplify the size of the scalar metric perturbations. In the following we will show that the entropy mode studied in this paper can provide an efficient mechanism to enhance the spectrum of curvature fluctuations.

The basic idea is as follows: We expect that the entropy field will rapidly grow during the final instances of the contracting phase. In this case, isocurvature perturbations can be converted into curvature perturbations as described by (\ref{dotzeta1}) \footnote{As we discuss in our companion paper \cite{inprep}, they can also be parametrically amplified when they pass through the bouncing phase.}. After the bounce, the evolution of the universe will enter the normal thermal history. In contrast to the bounce scenario without entropy modes, in the current model the energy scale for the bounce is able to be lowered and so the amplitude of tensor spectrum can be suppressed. Therefore, our model can solve the problem of large tensor-to-scalar ratio which
generally exists in nonsingular bounce cosmologies \cite{Wands:2008tv, Cai:2008qw}. Moreover, if the energy scale can be lowered to the TeV level, our scenario may lead to a window to test a bounce in particle physics.

\section{Evolution of Isocurvature Perturbations through the Bounce}

As the universe contracts in the matter-dominated phase, the Hubble parameter increases. When it catches up to the mass $m$ of the background field $\phi$, the
oscillations of this field will freeze out. After freeze-out, $\phi$ will slowly climb up the potential. As long as the total energy density is dominated by $\phi$, the
equation of state will decrease from $w=0$ to $w\simeq-1$. Therefore,  the universe will enter a contracting deflationary period and then approach the bouncing point. During this phase, both the isocurvature field $\chi$ and the ghost field $\psi_g$ will increase exponentially in amplitude. Thus, after a brief deflationary phase either $\chi$ or $\psi_g$ will come to dominate. If the energy density of $\psi_g$ is larger, then the bounce point is reached when the energy densities of $\phi$ and $\psi_g$ become equal. We will denote this case as ``Case 2''. In the other case (``Case 1'') there will be another brief phase of contraction driven by $\chi$, followed by another brief deflationary phase and a bounce when $\rho_{\chi} = |\rho_{\psi_g}|$. Note that the deflationary phases can be understood as the time reversal of a period of
slow-roll inflation.

\subsection{Deflationary period}

We first study the evolution of the entropy scalar $\chi$ in the deflationary period. From studies of chaotic type inflation models, it is well known that the background field $\phi$ freezes out when its amplitude $\tilde\phi$ becomes of the order of the
Planck mass, more specifically when
\begin{eqnarray}
 \tilde\phi \, \equiv \, \tilde\phi_m \, = \, \frac{2m_{pl}}{\sqrt{3\pi^3}} \, .
\end{eqnarray}
This leads to a nearly constant Hubble parameter
\begin{eqnarray}
 H_m \, = \, -\frac{4m}{3\pi}
\end{eqnarray}
and an almost exponentially decreasing scale factor for the universe. In this period, the scalars $\chi$ and $\psi_g$ will grow exponentially and the density in one of them will soon catch up to the density of background field. The background equation
of motion (\ref{eomchi}) yields
\begin{eqnarray}\label{chid}
 \chi \, \simeq \, \chi_m\exp\bigg[\frac{4m}{\pi}(t-t_m)\bigg]~,
\end{eqnarray}
which evolves proportional to $a^{-3}$. Here, the time $t_m$ is the end of the matter-dominated epoch of contraction, and the subscripts ``m'' on the fields $\phi$ and $\chi$ indicate the values of those field at the time $t_m$.

We will write
\begin{eqnarray}
 \chi_m \, &\equiv& \, d_1 \phi_m \, \nonumber \\
 (\psi_g)_m \, &\equiv& \, d_2 \phi_m \, ,
\end{eqnarray}
where the values of the coefficients $d_1$ and $d_2$ (both much smaller than $1$) depend on the initial conditions. We will denote the larger of the two numbers $d_1$ and $d_2$ by $d$. The deflationary phase will end when the energy of one of the fields $\chi$ and $\psi_g$ catches up to that in $\phi$. This will occur at the moment $t_d$ given by
\begin{eqnarray}
t_d - t_m \, \simeq \, \frac{\pi}{4m}\ln\frac{\pi}{4 d_1}
\end{eqnarray}
in Case 1 and
\begin{eqnarray}
t_d - t_m \, \simeq \, \frac{\pi}{2m}\ln\frac{m}{d_2 M}
\end{eqnarray}
in Case 2. The difference between the two cases is due to the fact that $\psi_g$ is oscillating in the deflationary phase, but $\chi$ is not. Hence, the expressions for the energy density of the two fields take a different form.

Let us now turn to the evolution of the entropy fluctuations which is given by Eq. (\ref{eqv}). In a de Sitter phase of contraction the dominant solution is
\begin{eqnarray}
v_k \, \sim \, a^{-2}
\end{eqnarray}
which corresponds to
\begin{eqnarray}
\delta \chi_k \, \sim \, a^{-3} \, .
\end{eqnarray}
Hence, it follows that the power spectrum of $\delta \chi$ gets enhanced during the deflationary phase by a factor
\begin{eqnarray}
 \label{amplif}
{\cal C} \, = \, (\frac{\pi}{4d_1})^2
\end{eqnarray}
in Case 1 and
\begin{eqnarray}
 \label{amplif2}
{\cal C} \, = \, (\frac{m}{d_2 M})^4
\end{eqnarray}
in Case 2.

Note, however, that this amplification does {\it not} lead to a suppression of the tensor to scalar ratio since the tensor modes at the same rate during the deflationary phase. It is only due to matter couplings that a difference between scalar and tensor modes will result. Below, we will see that such an amplification mechanism affecting only the scalar modes results in the post-deflationary phase, a phase driven by the kinetics of the fields.

We do need to worry about the absolute magnitude of the fluctuations. There is the danger that the amplification in the deflationary phase cause the perturbations become nonlinear. Then, we would lose perturbative control of the analysis. Presumably, this would also imply that space fragments into a gas of black holes rather than undergoing a smooth bounce. However, from (\ref{amplif}) and from  the fact that the fluctuations start out as quantum vacuum perturbations, it appears that this constraint on the parameters is a fairly mild one.

\subsection{Passing through the bouncing phase}

In this subsection we will give a rough analysis of how the fluctuations in $\chi$ pass through the bouncing phase. We will discover that the coupling between the matter fields due to the kinetic terms results in a growth of the entropy fluctuations without corresponding growth of the tensor modes. We call this process {\it kinetic amplification}.

After the period of deflation, the scalar $\chi$ continues to roll fast, and may even dominate over the background scalar $\phi$ (in Case 1 - see the beginning of this section for the definition of what the Cases are). In this case, the universe evolves with a background equation of state which varies from $w \simeq -1$ during deflation to $w \simeq 1$ during $\chi$ domination. In this case, the dominant term in the background Friedmann equation is the kinetic term of the scalar $\chi$, and leads to a nearly conserved quantity
\begin{eqnarray}\label{chiH}
\dot\chi/H \, \simeq \, \dot\chi_d/H_d \, \simeq \, \dot\chi_d/H_m~,
\end{eqnarray}
during this short period. In addition, from Eq. (\ref{chid}) we see that
\begin{eqnarray}\label{dchid}
 \dot\chi_d \, = \, \frac{4m}{\pi}\chi_d \, \simeq \, \frac{4m}{\pi}\tilde\phi_m~.
\end{eqnarray}

As discussed earlier, eventually the energy density in the ghost field $\psi_g$ catches up and the bounce phase will be triggered. Thus, the universe will exit the contraction phase and the equation of state of the universe will cross $-1$ and fall to negative infinity. In order to study the evolution of perturbations through the bouncing phase analytically, it is convenient to model the evolution of the Hubble parameter in the
following form:
\begin{eqnarray}\label{Hbounce}
H \, = \, \alpha t~,
\end{eqnarray}
with some positive coefficient $\alpha$ which has dimensions of $k^2$ and whose magnitude is set by the microphysics of the bounce. For example, $\alpha\simeq m^2$ in the model of Lee-Wick Cosmology\cite{Cai:2008qw}, while in Loop Quantum Cosmology
$\alpha$ is determined by a regularized Planck mass\cite{Bojowald:2008zzb, Mielczarek:2009zw}. Note that in the above parametrization we have chosen the time of the bounce to be zero.

To study the perturbation of the $\chi$ field, it is necessary to go back to the complete form of the perturbation equations given by (\ref{pertQ}). We will focus on the evolution of $Q_{\chi}$. From (\ref{pertQ}) we see that there is a mixing with $Q_{\phi}$. However, since the frequency of $\delta\phi$ is dominated by its mass $m$, $\delta \phi$ keeps oscillating even outside the Hubble radius. Hence, its amplitude is much smaller than that of $\delta\chi$ and therefore we neglect its contribution in the equation of motion for $\delta\chi$. Thus, $Q_{\chi}$ effectively decouples from $Q_{\phi}$. Focusing now on the mass term in the equation of motion for $Q_{\chi}$, we notice that since $\chi$ evolves as a stiff fluid with an effective equation of state $w=1$, the contribution of its kinetic term in the perturbation equation must be taken
into account. This contribution is given by the following term:
\begin{eqnarray}
 -\frac{8\pi}{a^3m_{pl}^2}\bigg(\frac{a^3\dot\chi^2}{H}\bigg)^{.} \,
 \simeq \, -\frac{72\pi\tilde\phi_m^2}{m_{pl}^2}(\alpha+3\alpha^2t^2)~,
\end{eqnarray}
where we have used the relations (\ref{chiH}), (\ref{dchid}) and (\ref{Hbounce}).

Consequently, we obtain the following equation of motion for the Fourier modes of the gauge invariant perturbation variable $Q_{\chi}$,
\begin{eqnarray}\label{pertQb}
 \ddot{Q}_{\chi} + 3H\dot{Q}_{\chi} + \frac{k^2}{a^2}Q_{\chi}
 + m_{\chi,{\rm eff}}^2Q_{\chi} \, \simeq \, 0 \, .
\end{eqnarray}
In the bouncing phase the effective mass term is given by
\begin{eqnarray}
m_{\chi,{\rm eff}}^2 \, &=& \, g^2\tilde\phi_m^2 -
\frac{72\pi\alpha\tilde\phi_m^2}{m_{pl}^2}(1+3\alpha{t}^2) \nonumber \\
&=&\, \frac{4g^2m_{pl}^2}{3\pi^3} - \frac{96\alpha}{\pi^2}(1+3\alpha{t}^2)~.
\end{eqnarray}
Thus, we find that in the bouncing phase the contribution of the kinetic term to the perturbations of the $\chi$ field yields a tachyonic mass term. This results in an increase in the magnitude of the fluctuations, a process which we call {\it kinetic amplification}. Note that this amplification does not affect the tensor modes.

In analogy with the theory of preheating (see e.g. \cite{Allahverdi:2010xz} for a recent review), we suggest that the investigation of the perturbation $Q_\chi$ can be simplified by redefining the fluctuation variable
\begin{eqnarray}
 \delta{X}_k \, \equiv \, a^{3/2}{Q_\chi}_k \, .
\end{eqnarray}
Then, we obtain a simpler form of the equation of motion for the fluctuations. Instead of Eq. (\ref{pertQb}) we now have
\begin{eqnarray}\label{eqX}
 \delta\ddot X_k + \omega_k^2\delta X_k \, = \, 0~,
\end{eqnarray}
where
\begin{eqnarray}\label{wk2}
 \omega_k^2 \, \equiv \, \frac{k^2}{a^2} + m_{\chi,{\rm eff}}^2
 + \frac{3}{4}H^2 - \frac{3}{2}\dot{H}~,
\end{eqnarray}
is the frequency for the k'th mode of the isocurvature perturbations.

Consequently, if
\begin{eqnarray}
 \alpha \, \gg \, g^2m_{pl}^2 \, ,
\end{eqnarray}
there exists a period $-t_b  <  t  <  t_b$ with
\begin{eqnarray}
 t_b \, \simeq \, |H_m|/\alpha  =  4m/3\alpha\pi
\end{eqnarray}
near the bounce point during which the adiabaticity condition is violated and the dispersion relation of the $\chi$ perturbation is unstable in the infrared limit. In this case the evolution of the $\chi$ perturbation can be approximately expressed as \cite{Enqvist:2004ey}
\begin{eqnarray}
 \delta X \, \propto \, \exp[|\int dt \omega(t)|]
\end{eqnarray}
in which the integral runs from $-t_b$ to $t_b$ if $\alpha\gg g^2m_{pl}^2$.

As a consequence, we find that the amplitude of the entropy fluctuations obtains a tachyonic amplification around the bounce point. Since this instability arises from the kinetic term of the entropy field, we call this process {\it kinetic amplification}. One can describe the amplification of the infrared modes of isocurvature perturbations in this resonance stage as follows:
\begin{eqnarray}\label{dchib}
 \delta\chi_b \, = \, {\cal F} \delta\chi_d ~,
\end{eqnarray}
with an amplification factor
\begin{eqnarray}\label{Ffactor}
 {\cal F} \, &=& \, \exp[|\int_{-tb}^{tb} dt w(t)|] \nonumber\\
 &\simeq& \, e^{\sqrt{y(2+y)} + \frac{3}{\sqrt{2}}\sinh^{-1}(\frac{2\sqrt{y}}{3})}~.
\end{eqnarray}
In the above calculation, we have introduced a variable $y=\frac{m^2}{\alpha}$ which characterizes how fast the bouncing phase occurs.

\begin{figure}[htbp]
\includegraphics[scale=0.8]{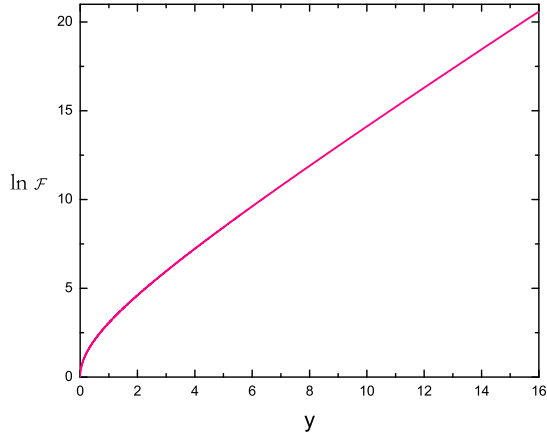}
\caption{Plot of the exponent in the amplification factor $\ln{\cal F}$ as a function of the variable $y$ in the bouncing phase. The permitted range of $y$ runs from $0$ to
$\frac{m^2}{g^2m_{pl}^2}$, and we specifically choose $g^2m_{pl}^2=m^2/16$. } \label{fig:Ffactor}
\end{figure}

To understand the exponent appearing in the amplification factor ${\cal F}$, we plot its evolution as a function of the variable $y$ in Fig. \ref{fig:Ffactor}. One can see that in the limit of a fast bounce $y \rightarrow 0$ (which corresponds to $\alpha \gg m^2$) the contribution of kinetic amplification is quite limited since the duration of the bouncing phase is very short. In the case of the Lee-Wick cosmology $\alpha \simeq m^2$ and thus gives $y \simeq1$, and correspondingly we obtain a value of the amplification factor ${\cal F} \simeq 21.3$ which is quite sizable.

One interesting issue which should be noticed that, if we choose the value of the coefficient $\alpha$ to be between $g^2m_{pl}^2$ and $m^2$, the amplitude of the isocurvature perturbation will be distinctly amplified. This case corresponds to a slow bouncing phase. This amplification becomes more efficient if we take the coupling
parameter $g^2$ to be slightly negative which provides a flat tachyonic potential for the entropy field.

\subsection{The $\chi$ power spectrum in the expanding phase}

After the bounce, the universe will eventually re-emerge into a matter-dominated phase, this time a phase of cosmological expansion. Since the dominant mode of both scalar and tensor fluctuations on super-Hubble scales is constant, no further amplification of the entropy fluctuations results, nor do scalar and tensor perturbations evolve asymmetrically.

Now let us take into account both of the amplification effects on the $\chi$ fluctuations discussed in the two previous subsections. Combining the relations
(\ref{amplif}) and (\ref{dchib}) and applying the relation (\ref{dchic}) at the beginning of the period of deflation at the time $t_{m}$, we eventually obtain the following amplitude of the isocurvature perturbations after the bounce
\begin{eqnarray}
 \delta\chi_f \, \simeq \, \frac{m}{3\pi^2}{\cal C}^{\frac{1}{2}} {\cal F}~,
\end{eqnarray}
on super-Hubble scales.

\section{Generation of metric perturbations}

In the above section we have studied the dynamics of isocurvature perturbations in a particular implementation of the matter bounce scenario. Next, we will study the induced curvature fluctuations, the ``matter bounce curvaton'' mechanism. The transfer takes place in the deflationary phase and in the bouncing phase when the equation of state of the background is changing because of the contribution of the entropy field.

\subsection{Linear fluctuations}

We first study the linear curvature fluctuations induced by the isocurvature fluctuation. To do this, we return to the general formula (\ref{dotzeta}) which describes the transfer of isocurvature to curvature fluctuations. Making use of the fact that
\begin{eqnarray}
{\dot H} \, = \, -\frac{4\pi}{m_{pl}^2} \sum_i {\dot{\phi_i^2}}~,
\end{eqnarray}
and inserting the definition of $Q_i$ we find that on large scales (\ref{dotzeta}) reduces to
\begin{eqnarray} \label{dotzeta2}
 {\dot \zeta} \, = \, - 4\pi m_{pl}^{-2} \sum_i \frac{H \delta \phi_i}{\dot{\phi_i}} (\frac{\dot\phi_i^2}{\dot{H}})^{.}~.
\end{eqnarray}

Since the spectrum of the ghost field $\psi_g$ has a blue spectrum due to the large mass of the ghost, the ghost field contribution to (\ref{dotzeta2}) can be neglected, and only two terms remain in the sum, the one due to $\phi$ and the one due to the entropy field. Since in our realization of the bounce scenario the regular matter field $\phi$ is massive, its contribution to (\ref{dotzeta2}) is also negligible. A second way to see that the term proportional to $\delta \phi$ gives a negligible contribution to (\ref{dotzeta2}) is to realize that until the entropy field $\chi$ begins to dominate the background, the factor ${\dot \phi}^2 / {\dot H}$ is approximately constant and thus the coefficient of the $\delta \phi$ term vanishes. Thus, (\ref{dotzeta2}) can be approximated by
\begin{eqnarray}
 {\dot \zeta} \, \simeq \, - 4\pi m_{pl}^{-2} \frac{H \delta \chi}{\dot{\chi}}
 (\frac{\dot\chi^2}{\dot{H}})^{.}~.
\end{eqnarray}
The right hand side of this expression vanishes at the beginning of the deflationary phase and gradually grows until the bounce phase is reached. We can approximately integrate the equation to obtain
\begin{eqnarray} \label{dotzeta3}
 \Delta \zeta \, \simeq \, - 4\pi m_{pl}^{-2} \frac{H}{{\dot H}} {\dot \chi} \delta \chi \, .
\end{eqnarray}

Let us first evaluate the magnitude of the induced curvature fluctuations in Case 1,
when $\rho_{\chi} > |\rho_{\psi_g}|$. In this case the contribution of the ghost condensate field to ${\dot H}$ at the time when the deflationary period ends can be neglected and
\begin{eqnarray}
 {\dot H} \, \simeq \, - 4\pi m_{pl}^{-2} \bigl(\dot\phi^2 +\dot\chi^2\bigr) \, .
\end{eqnarray}
When the $\chi$ field starts to dominate the energy density, the magnitudes of ${\dot \chi}$ and ${\dot \phi}$ are comparable, and hence
\begin{eqnarray}
 \Delta \zeta \, \simeq \, \frac{H}{{\dot \chi}} \delta \chi \, .
\end{eqnarray}
Making use of the background solution for the scalar $\chi$ and the amplification of the amplitude of the isocurvature perturbation $\delta\chi_f$ before and during the bounce, we finally obtain the following estimate for the induced curvature perturbation
\begin{eqnarray}\label{zetaf}
 \Delta \zeta \, \simeq \, \frac{m}{6\sqrt{3\pi}m_{pl}} {\cal C}^{\frac{1}{2}} {\cal F}~,
\end{eqnarray}
where we have used relation (\ref{chiH}) which is approximately valid during the deflationary and bounce periods.

Next, let us turn to Case 2. Now, ${\dot H}$ is dominated by ${\dot \psi_g}$ instead of by ${\dot \chi}$ in the bounce region. The numerator in (\ref{dotzeta3}) remains the
same as for Case 1. Hence, the overall amplitude of the induced curvature fluctuations is suppressed by a factor ${\cal S}$, where
\begin{eqnarray}
 {\cal S} \, = \, (\frac{{\dot \chi}}{{\dot \psi_g}})^2 \, ,
\end{eqnarray}
compared to what happens in Case 1.

The tensor modes are also amplified during the deflationary phase, as mentioned earlier. Thus, they also obtain the amplification factor ${\cal C}$. Their overall amplitude is hence the same as that of the induced curvature fluctuations, except for
the factor ${\cal F}$ due to kinetic amplification. Hence, the predicted tensor to scalar ratio $r$ in Case 1 is
\begin{eqnarray} \label{tensor}
 r \, \simeq \, {\cal F}^{-2} \, .
\end{eqnarray}
Its magnitude is determined by two parameters, namely the mass of the heavy field $m$ and the slope of the bouncing phase $\alpha$.

According to the latest observations from the Wilkinson Microwave Anisotropy Probe 7-year data (WMAP7) \cite{Komatsu:2010fb}, the amplitude of curvature perturbation is $\zeta \simeq 4.92\times10^{-5}$ and the spectral tilt is $|n_{\chi}|\lesssim 0.03$. Thus, the observed amplitude of the curvature perturbations can be obtained in our scenario if we require
\begin{eqnarray}
 m \simeq 9.02\times 10^{-4}{\cal F}^{-1} m_{pl} \, .
\end{eqnarray}

For completeness, let us mention (as was already used above in (\ref{tensor})) that in our model the amplitude of tensor perturbation is about $h\sim H/m_p$. Specifically,
following the calculations in Ref. \cite{Cai:2008qw}, one obtains the following primordial power spectrum for the tensor perturbations at the end of the matter dominated phase of contraction
\begin{eqnarray}
 P_T(k)\, \equiv \, \frac{32k^3}{\pi m_{pl}^2}|h_k|^2
 \, = \, \frac{16 H^2}{9 \pi m_{pl}^2}~,
\end{eqnarray}
During the deflationary phase, there is an enhancement by a factor of ${\cal C}$. The tensor-to-scalar ratio is given by
\begin{eqnarray}
 r \, \simeq \, 35{\cal F}^{-2}~.
\end{eqnarray}
This ratio can be greatly suppressed by a large value of the factor ${\cal F}$. In the
specific model of Lee-Wick cosmology, $\alpha \simeq m^2$ so that $r \simeq 0.08$. With the WMAP7 data combined with BAO and SN, the current limit on the tensor-to-scalar ratio is given by $r < 0.2$.

Our analytical calculations involve some approximations. Specifically, in the deflationary phase the Hubble parameter is not exactly a constant and the background scalar $\phi$ is still rolling slowly, but in our analytical analysis we have taken their values at the freeze-out time. As a consequence, it is important to confirm the results by a numerical integration of the full equations.

\begin{figure}[htbp]
\includegraphics[scale=0.8]{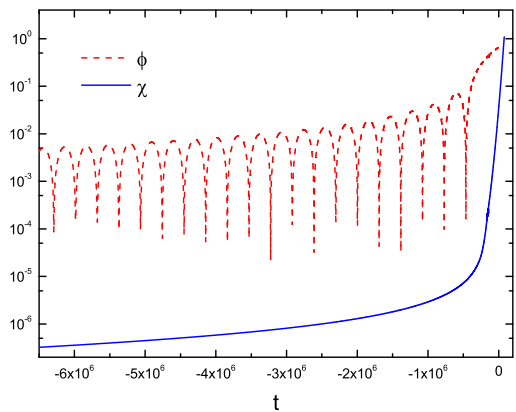}
\caption{Evolution of the background fields $\phi$ and $\chi$ before the bounce as a function of cosmic time (horizontal axis). The background fields are plotted in dimensionless units by normalizing by $m_{pl}$. Similarly, the time axis is displayed in units of $m_{pl}^{-1}$. The parameters $m$ and $g$ appearing in the potential were chosen to be $m = 10^{-5} {m_{pl}}$, $g = 2.5 \times 10^{-6}$, and we assume initially $d = 10^{-4}$. The initial conditions for $\phi$ were taken to be $\phi_i= 1.19 \times 10^{-5} {m_{pl}}, \,\, \dot\phi_i = - 1.24 \times 10^{-9} {m_{pl}}^2$.} \label{fig:bg}
\end{figure}

We first numerically study the evolution of two scalars. As demonstrated in Fig. \ref{fig:bg}, in the era of contraction, the heavy field $\phi$ keeps oscillating with an increasing amplitude and dominates the background evolution, while the light field $\chi$ tracks the evolution of $\phi$. When the amplitude of $\phi$ grows up to of Planck mass, $\chi$ rolls rapidly and finally catch up to $\phi$ near by the bounce. This numerical result is consistent with the analytic estimates appeared in Eqs. (\ref{chi}), (\ref{chid}). However, in our numerical calculations the background evolution ceases around the moment of $\chi$ domination which is before the big bounce. Since in the current paper we wish to discuss the dynamics of isocurvature perturbations in a general case, we did not concentrate on a detailed mechanism leading to a bouncing phase but take a parametrization of the Hubble parameter $H = \alpha t$ instead.

\begin{figure}[htbp]
\includegraphics[scale=0.8]{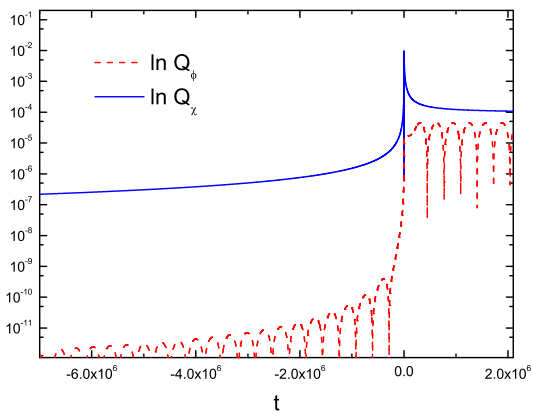}
\caption{Evolution of the gauge-invariant scalar perturbations $Q_{\phi}$ and $Q_{\chi}$ for super-Hubble mode $k=2\times10^{-6}$ in our model. The horizontal axis is cosmic time.  The scalar perturbations are plotted in dimensionless units by normalizing by $m_{pl}$. In the parametrization function $\alpha=10^{-6}m_{pl}^2$
is chosen. The background parameters are the same as in Fig. \ref{fig:bg}.} \label{fig:perts}
\end{figure}

Based on the background solution obtained above, we then solve the equations of motion (\ref{pertQ}) numerically. We show the evolution of the gauge-invariant scalar perturbation variables $Q_{\phi}$ and $Q_{\chi}$ in Fig. \ref{fig:perts}. In the
numerical calculations, we consider these perturbation modes to originate from vacuum fluctuations inside the Hubble radius \footnote{Numerical analyses of inflationary preheating were performed with a massive scalar \cite{Bassett:1998wg, Bassett:1999mt, Finelli:1998bu} and a massless scalar \cite{Bassett:1999cg, Finelli:2000ya} (see also \cite{Taruya:1997iv}) respectively.}. One can see from Fig. \ref{fig:perts} that the perturbation of the heavy field (the adiabatic mode after the bounce) catches up to the amplitude of the entropy field perturbation. This demonstrates the effectiveness of the ``matter bounce curvaton'' mechanism discussed here. Note that $Q_{\phi}$ keeps oscillating through the whole evolution. The amplitude of the fluctuation $Q_\chi$ grows before the bounce, however, with a different slope than the growth of $Q_{\phi}$. This difference in slope is due to the transfer of isocurvature to curvature fluctuation. Overall, the numerical results are consistent with the analytic calculations.

\subsection{Local non-Gaussian fluctuations}

Next we take a first look at an important byproduct of the ``matter bounce curvaton'' scenario, namely a contribution to the induced non-Gaussian distribution of metric fluctuations which will be of local form.

There are two important periods for generating a large non-Gaussianity in a bouncing universe. The first is the evolution of metric perturbations on super-Hubble scales in the contracting phase which at leading non-linear order generates non-Gaussianities as studied in Refs. \cite{Cai:2009fn, Cai:2009rd}. This generates an effective value of the non-linearity parameter $f_{nl}$ which is of order one, the largeness being due to the fact that what corresponds to the inflationary slow-roll parameter is now a constant of order one. A distinctive feature of this source of non-linearity is a particular shape of the bispectrum. The shape is different than the one which appears
in simple inflationary models since in the contracting phase $\zeta$ is not constant and hence terms depending on ${\dot \zeta}$ which vanish in inflationary cosmology contribute in an important way to the bispectrum.

The second effect is a contribution to the bispectrum of local form which results from the curvaton mechanism. It is due to the transformation of entropy perturbations into curvature fluctuations which happens around the bounce point. Here, we focus on the second effect.

A naive way to investigate the non-Gaussianity is to study the cosmological fluctuations to second order\cite{Acquaviva:2002ud} \footnote{The issue of inflationary non-Gaussianities has been studied in detail e.g. in Refs. \cite{Maldacena:2002vr, Chen:2006nt} and comprehensively reviewed by \cite{Bartolo:2004if}. Moreover, the non-Gaussian fluctuations during inflationary preheating have been analyzed in \cite{Enqvist:2004ey}; see also \cite{Vernizzi:2006ve, Jokinen:2005by, Barnaby:2006cq}.}. Recalling the time evolution of the curvature perturbation given in Eq. (\ref{dotzeta1}) and taking the next-to-leading order corrections into account, we are able to decompose the curvature perturbation into a Gaussian part and a non-linear one, where the second part evolves according to the following equation\cite{Langlois:2006vv, Koyama,EkpNG,Lehners:2007wc}:
\begin{eqnarray}
 \dot\zeta_{NL} \, = \,
 \frac{H}{\dot\sigma^2}(V_{,ss}+4\dot\theta^2)\delta{s}^2 - \frac{H}{\dot\sigma^3}V_{,\sigma}\delta{s}\dot\delta{s}~,
\end{eqnarray}
on long wavelengths. In the above formula, we have introduced the adabatic field $\sigma=\int(\dot\phi^2+\dot\chi^2)^{\frac{1}{2}}dt$, the isocurvature field $s=(\dot\phi\chi-\dot\chi\phi)/\sigma$, and the trajectory angle $\theta$ satisfying $\cos\theta=\dot\phi/\dot\sigma$ and $\sin\theta=\dot\chi/\dot\sigma$.

As usual, we define a nonlinearity parameter $f_{NL}$ to measure the amplitude of non-Gaussian perturbations. This parameter is defined by the relation
\begin{eqnarray}
 \zeta \, = \, \zeta_{L} + \frac{3}{5}f_{NL}\zeta_{L}^2 \, ,
\end{eqnarray}
where $\zeta_{L}$ is the linear order value of curvature perturbation. As analyzed
in the previous section, we consider the leading non-Gaussianity when modes are at super-Hubble scales, and thus the non-Gaussianity is of the local type\cite{Maldacena:2002vr, Babich:2004gb}.

We follow the time evolution of the curvature perturbations and use the solution (\ref{dchic}), and then obtain the non-linear part of the curvature perturbation,
\begin{eqnarray}
 \zeta_{NL} &\simeq& \int_m^f H\frac{m^2\dot\phi^2}{\dot\sigma^6} \bigg[\dot\chi^2(1+\frac{4m^2\phi^2}{\dot\sigma^2})\delta\chi^2 -\phi\dot\phi\delta\chi\delta\dot\chi\bigg]dt \nonumber\\
 &\simeq& \frac{32m}{\pi m_{pl}^3}(1-\frac{8d^2{\cal C}m_{pl}}{\pi^2m})\delta\chi_f^2~,
\end{eqnarray}
where the main contribution is provided by the integral from $t_m$ to $t_d$ during which the scalar $\chi$ comes to dominate. Moreover, combined with the linear form of the curvature perturbation (\ref{zetaf}), we obtain the leading term of the nonlinearity parameter
\begin{eqnarray}
 f_{NL} \, \simeq \, -\frac{5120}{\pi^6}d^2{\cal C} \, .
\end{eqnarray}

As discussed in the previous sections, there are two cases for the choice of $d$ and ${\cal C}$ parameters depending on how the scalar field $\chi$ becomes dominant. For Case 1, we get
\begin{eqnarray}
 f_{NL} \, \simeq \, -3.3~,
\end{eqnarray}
by making use of ${\cal C}=(\pi/4d)^2$; for Case 2, we obtain
\begin{eqnarray}
f_{NL}\, \simeq \, -5.3\frac{m^4}{d^2M^4}~.
\end{eqnarray} 
In both two cases, we see that it is possible to generate a sizable non-Gaussianity in our model by this mechanism. Specifically, a large non-Gaussianity of local type can be obtained for a small value of $d$ in Case 2. For example, taking the value of the parameters to be such that $M=100m$ which satisfies the theoretical bound discussed at the end of the Appendix, we find that with $d=10^{-2}\frac{m}{M}$ one gets $f_{NL} \simeq -5.3$, a value which is consistent with current observational constraints which give $-36<f_{NL}<100$ at the $2\sigma$ level, but a value which is not too small to be detectable in upcoming experiments.

Although the mechanism discussed above is quite efficient to generate a sizable non-Gaussianity, there are other sources which could potentially also give a large value
of $f_{NL}$. For example, we expect that preheating effects will lead to a major contribution. In particular, they will affect the shape of the non-Gaussianities, a topic we did not study here. We will leave the detailed study of this issue to a
future publication.

\section{Conclusions and Discussion}

In this paper we introduced the ``matter bounce curvaton'' scenario, a new mechanism to generate a scale-invariant primordial spectrum of cosmological perturbations in the framework of a nonsingular bounce cosmology with a matter-dominated phase of contraction. This model is able to solve the ``large $r$ problem'' from which many bouncing cosmologies suffer, while preserving the scale-invariance of the power spectrum. Our scenario is based on taking into account the contributions of light fields to the curvature fluctuations. The light fields act as curvatons. They acquire a scale-invariant spectrum of fluctuations on super-Hubble scales during the contracting phase. As we have shown, the change in the equation of state immediately before and during the bounce phase leads to the transfer of the fluctuations to the adiabatic mode, maintaining the scale-invariance of the spectrum.

We have studied our mechanism in a concrete realization of the matter bounce scenario, namely the quintom bounce, a scenario in which the violation of the null energy condition which is required to obtain a non-singular bounce is obtained via the addition of a ghost field. However, the mechanism will also be effective in other realizations of the matter bounce.

As we have shown here, just before and during the bounce, the entropy fluctuations are amplified by an effective tachyonic mass due to a kinetic resonance. We have called this process ``kinetic amplification''. Whereas the scalar modes are amplified by this mechanism, the gravitational waves are not. Thus, our scenario leads to a suppression of the tensor to scalar ratio $r$. The kinetic amplification mechanism is more effective if the bounce is slower, since then the time interval during which the entropy fluctuations see an effective tachyonic mass is longer. The kinetic amplification mechanism is also more effective if the energy density in the entropy field initially dominates over the absolute value of the energy density in the ghost field.

We also have taken a first look at non-Gaussianities from our mechanism. Specifically, we have discussed one source of local type non-Gaussianity generated in our curvaton mechanism. We find, for typical parameter values, the predicted amplitude is quite sizable to be observationally relevant. There are other sources, however, which are expected to lead to even larger effects. Note that there have recently been some attempts to seed sizable non-Gaussianity in the framework of inflationary preheating\cite{Dvali:2003em, Barnaby:2006km, Chambers:2007se, Huang:2009vk}. The distinguishing difference between their models and ours is that in our case the non-Gaussian fluctuations arise even for modes in the far infrared, modes relevant for current cosmological observations. However, those in inflation are contributed by a narrow resonance which puts an upper bound on their wavelength \cite{Byrnes:2008zz}.

\begin{acknowledgments}

Y.F.C. is supported by funds from the Department of Physics at Arizona State University. Y.F.C thanks McGill University for hospitality while this work was finalized. The research of R.B. is supported in part by an NSERC Discovery Grant at McGill University and by funds from the Canada Research Chairs program. The research of X.Z. is supported in part by the National Natural Science Foundation of China under Grants No. 10975142, 10821063 and 11033005, by the 973 program No. 1J2007CB81540002, and by the Chinese Academy of Sciences under Grant No. KJCX3-SYW-N2.

\end{acknowledgments}

\section{Appendices}

The the first part of this Appendix we study the evolution equation for $\zeta$, in the second we look at the evolution of the entropy fluctuations in the deflationary phase.

\subsection{General Analysis of $\zeta$}

As in the main text, we assume that matter is described by a collection of scalar fields. In this case, at linear order, the generic form of the equation for the time derivative of $\zeta$ is
\begin{eqnarray}\label{dotzeta10}
 \dot\zeta \, = \, -\frac{H}{\dot{H}}\nabla^2\Phi -
 4\pi GH\sum_i\frac{Q_i}{\dot\phi_i}(\frac{\dot\phi_i^2}{\dot{H}})^{.}~.
\end{eqnarray}
On large scales, the first term of the r.h.s of Eq. (\ref{dotzeta10}) is negligible.
The second term describes the transfer of entropy to adiabatic fluctuations, the term we are interested in.

At first, let us see how the above multi-field formula applies to the double field case which is usually analyzed in the literature. For a model with two fields $\phi_1$ and $\phi_2$, we get
\begin{eqnarray}
 \dot\zeta
 = -\frac{H}{\dot{H}}\nabla^2\Phi
 +H\frac{Q_1}{\dot\phi_1}(\frac{\dot\phi_1^2}{\dot\phi_1^2+\dot\phi_2^2})^{.} +H\frac{Q_2}{\dot\phi_2}(\frac{\dot\phi_2^2}{\dot\phi_1^2+\dot\phi_2^2})^{.}~.
\end{eqnarray}
One notices that
\begin{eqnarray}
 (\frac{\dot\phi_1^2}{\dot\phi_1^2+\dot\phi_2^2})^{.} +(\frac{\dot\phi_2^2}{\dot\phi_1^2+\dot\phi_2^2})^{.} = 0~,
\end{eqnarray}
and thus one obtains
\begin{eqnarray}
 \dot\zeta
 &=& -\frac{H}{\dot{H}}\nabla^2\Phi
  +H\frac{Q_1}{\dot\phi_1} [\frac{1}{2}(\frac{\dot\phi_1^2}{\dot\phi_1^2+\dot\phi_2^2})^{.} -\frac{1}{2}(\frac{\dot\phi_2^2}{\dot\phi_1^2+\dot\phi_2^2})^{.}]  \nonumber \\
  & & + H\frac{Q_2}{\dot\phi_2} [\frac{1}{2}(\frac{\dot\phi_2^2}{\dot\phi_1^2+\dot\phi_2^2})^{.} -\frac{1}{2}(\frac{\dot\phi_1^2}{\dot\phi_1^2+\dot\phi_2^2})^{.}] \nonumber\\
 &=& -\frac{H}{\dot{H}}\nabla^2\Phi
  + \frac{H}{2} (\frac{Q_1}{\dot\phi_1}-\frac{Q_2}{\dot\phi_2}) (\frac{\dot\phi_1^2-\dot\phi_2^2}{\dot\phi_1^2+\dot\phi_2^2})^{.}~,
\end{eqnarray}
which is exactly what is obtained in the usual literature.

In the following, we provide the detailed derivation of the equation (\ref{dotzeta10}) for the time derivative of $\zeta$ in the multiple field model. We start with the standard perturbed Einstein equations,
\begin{eqnarray}
 \ddot\Phi+4H\dot\Phi+(2\dot{H}+3H^2)\Phi &=& 4\pi G\delta{p}~,\\
 \dot\Phi+H\Phi &=& 4\pi G\delta{q}~,\\
 -3H(\dot\Phi+H\Phi)+\nabla^2\Phi &=& 4\pi G\delta\rho~,
\end{eqnarray}
where
\begin{eqnarray}
 \delta{p}  &=& \sum_i[\dot\phi_i(\delta\dot\phi_i-\dot\phi_i\Phi)-V_{,i}\delta\phi_i]~,\\
 \delta{q}  &=& \sum_i\dot\phi_i\delta\phi_i~,\\
 \delta\rho &=& \sum_i[\dot\phi_i(\delta\dot\phi_i-\dot\phi_i\Phi)+V_{,i}\delta\phi_i]~.
\end{eqnarray}

Recall that the gauge-invariant curvature perturbation variable $\zeta$ can be expressed as
\begin{eqnarray}
 \zeta = \Phi - \frac{H}{\dot{H}}(\dot\Phi+H\Phi)~.
\end{eqnarray}
Consequently, its time derivative is given by
\begin{eqnarray}
 \dot\zeta &=& \dot\Phi - (\dot\Phi+H\Phi) - \frac{H}{\dot{H}}[4\pi G\delta{p}-\dot{H}\Phi \nonumber \\
 & & -3H(\dot\Phi+H\Phi)] + \frac{H\ddot{H}}{\dot{H}^2}(\dot\Phi+H\Phi) \nonumber\\
 &=& -\frac{H}{\dot{H}}[4\pi G\delta{p}+4\pi G\delta\rho-\nabla^2\Phi] +\frac{H\ddot{H}}{2\dot{H}^2}\delta{q} \nonumber\\
 &=& -\frac{H}{\dot{H}} [8\pi G\sum_i\dot\phi_i\delta\dot\phi_i+2\dot{H}\Phi-2\nabla^2\Phi+\nabla^2\Phi] \nonumber \\
 & & +4\pi G\frac{H\ddot{H}}{\dot{H}^2}\sum_i\dot\phi_i\delta\phi_i \nonumber\\
 &=& -\frac{H}{\dot{H}}\nabla^2\Phi - 8\pi G\frac{H}{\dot{H}} [\sum_i\dot\phi_i\delta\dot\phi_i -
 \sum_i\dot\phi_i(\delta\dot\phi_i+3H\delta\phi_i) \nonumber \\
 & & -\sum_iV_{,i}\delta\phi_i]  + 4\pi G\frac{H\ddot{H}}{\dot{H}^2}\sum_i\dot\phi_i\delta\phi_i \nonumber\\
 &=& -\frac{H}{\dot{H}}\nabla^2\Phi - 8\pi G\frac{H}{\dot{H}}\sum_i\ddot\phi_i\delta\phi_i + 4\pi G\frac{H\ddot{H}}{\dot{H}^2}\sum_i\dot\phi_i\delta\phi_i \nonumber\\
 &=& -\frac{H}{\dot{H}}\nabla^2\Phi + 8\pi G H\sum_i\frac{\delta\phi_i}{\dot\phi_i} (-\frac{\dot\phi_i\ddot\phi_i}{\dot{H}}+\frac{\dot\phi_i^2\ddot{H}}{2\dot{H}^2}) \nonumber\\
 &=& -\frac{H}{\dot{H}}\nabla^2\Phi -4\pi GH\sum_i\frac{\delta\phi_i}{\dot\phi_i}(\frac{\dot\phi_i^2}{\dot{H}})^{.}~.
\end{eqnarray}
In the above calculation, we have used the background equation
\begin{eqnarray}
\dot{H} \, = \, -4\pi G\sum_i\dot\phi_i^2
\end{eqnarray}
a number of times. Moreover, it is convenient to introduce the Sasaki-Mukhanov
variables as defined in the Section II, $Q_i\equiv\delta\phi_i+\frac{\dot\phi_i}{H}\Phi$.
Then we precisely obtain Eq. (\ref{dotzeta}).

\subsection{Fluctuations in the Deflationary Phase}

Now we study the evolution of the entropy scalar $\chi$ in the deflationary process.
From studies of chaotic type inflation models, it is well known that the background
field $\phi$ freezes out when the amplitude $\tilde\phi$ becomes of the order
of the Planck mass, more specifically when $\tilde\phi_m=\frac{2m_{pl}}{\sqrt{3\pi^3}}$.
This leads to a nearly constant Hubble parameter $H_m=-\frac{4m}{3\pi}$ and an almost exponentially decreasing scalar factor for the universe. This phase is crucial to
achieving a nonsingular bounce, since the matter component, which effectively carries
the negative energy density (namely, the Lee-Wick partner field or a ghost condensate), will grow rapidly and soon catch up to the density of $\phi$.

Assuming that the Lee-Wick partner $\psi$ satisfies the following background equation of motion,
\begin{eqnarray}\label{EoMpsi}
 \ddot\psi + 3H\dot\psi + M^2\psi \, = 0~,
\end{eqnarray}
it is always oscillating around the stable point of its potential throughout the whole
evolution of the universe. Before the deflationary phase, both the amplitudes $\tilde\psi$ and $\tilde\phi$ are growing in the same rate. Therefore, it is convenient to define a ratio
\begin{eqnarray}
 d_2 \, \equiv \, \frac{\tilde\psi}{\tilde\phi}~,
\end{eqnarray}
which is almost constant until the moment $t_m$, the end of the matter-dominated phase of contraction.  Afterwards, solving the background equation of motion (\ref{EoMpsi}), one obtains
\begin{eqnarray}
 \tilde\psi \, \simeq \, \tilde\psi_m\exp\bigg[\frac{2m}{\pi}(t-t_m)\bigg]~,
\end{eqnarray}
which evolves proportional to $a^{-3/2}$. One can compute that the contribution of $\psi$ catches up to that of $\phi$ at the moment
\begin{eqnarray}
t_d \, = \, t_m+\frac{\pi}{2m}\ln\frac{m}{d_2M} \,
\end{eqnarray}
when $\tilde\psi_d \simeq \frac{2mm_{pl}}{9M}$.

We are able to compute the solution to the entropy field $\chi$ as well. This yields  $\chi\propto a^{-3}$. Then we continue to solve the perturbation equation in the infrared limit $k \rightarrow 0$, and obtain the amplitude of the isocurvature perturbation at the moment $t_d$ as follows,
\begin{eqnarray}
 \delta\chi_d &\simeq& \delta\chi_m\exp\bigg[\frac{4m}{\pi}(t_d-t_m)\bigg] \nonumber\\
 &\simeq& {\cal C}^{1/2}\delta\chi_m~,
\end{eqnarray}
where we have introduced an amplification factor ${\cal C}$, which is determined by
\begin{eqnarray}
 {\cal C} = (\frac{m}{d_2M})^4~,
\end{eqnarray}
indicating that the amplification on the isocurvature perturbation in deflationary process is depressed by the ratio $m/M$ but manifestly amplified by $d_2$. The coefficient $d_2$ is determined by the the ratio of two background scalar fields at low energy scale which is far away before the bouncing point. At that moment, the excitation of $\psi$ is strongly suppressed by the mass scale $M$ as illustrated in the Lee-Wick model and the ghost condensate, and thus $d_2<m/M$ is satisfied commonly.

Note that, the above analysis involves an important requirement that the coupling between the background scalar $\phi$ and the entropy field $\chi$ is enough week, i.e., $g<m/m_{pl}$. It is possible to choose a large value of the coupling constant which breaks this bound. This case corresponds to a stochastic resonance for the isocurvature perturbation before the bouncing phase, but would spoil the feature of scale-invariance as observed by current experiments. A detailed analysis of stochastic resonance on entropy field around the bounce will be performed in an accompanied paper in a following up project.

\end{document}